\documentclass[]{aastex631}

\begin{document}

\title[Technosignatures as anomalies]{Using anomaly detection to search for technosignatures in Breakthrough Listen observations}

\author{Snir Pardo}
\affiliation{School of Physics and Astronomy, Tel-Aviv University 
Tel-Aviv 69978, Israel. \\}
\author[0000-0003-1470-7173]{Dovi Poznanski}
\affiliation{School of Physics and Astronomy, Tel-Aviv University 
Tel-Aviv 69978, Israel. \\}
\affiliation{Cahill Center for Astrophysics, California Institute of Technology, Pasadena CA 91125, USA.\\}
\affiliation{Kavli Institute for Particle Astrophysics \& Cosmology, 452 Lomita Mall, Stanford University, Stanford, CA 94305, USA. \\}
\affiliation{Department of Physics, Stanford University, 382 Via Pueblo Mall, Stanford, CA 94305, USA. \\}
\author[0000-0003-4823-129X]{Steve Croft}
\affiliation{SETI Institute, 339 Bernardo Ave, Suite 200, Mountain View, CA 94043, USA. \\}
\affiliation{Breakthrough Listen, University of California, Berkeley, 
3501 Campbell Hall 3411, Berkeley, CA 94720, USA. \\}
\affiliation{Department of Physics, University of Oxford, Denys Wilkinson Building, Keble Road, Oxford, OX1 3RH, UK. \\}
\author[0000-0003-2828-7720]{Andrew P. V. Siemion}
\affiliation{SETI Institute, 339 Bernardo Ave, Suite 200, Mountain View, CA 94043, USA. \\}
\affiliation{Breakthrough Listen, University of California, Berkeley, 
3501 Campbell Hall 3411, Berkeley, CA 94720, USA. \\}
\affiliation{Department of Physics, University of Oxford, Denys Wilkinson Building, Keble Road, Oxford, OX1 3RH, UK. \\}
\author[0000-0002-7042-7566]{Matthew Lebofsky}
\affiliation{Breakthrough Listen, University of California, Berkeley, 
3501 Campbell Hall 3411, Berkeley, CA 94720, USA. \\}

\email{snir.pardo92@gmail.com,dovi@tau.ac.il}

\begin{abstract}

We implement a machine learning  algorithm to search for extra-terrestrial technosignatures in radio observations of several hundred nearby stars, obtained with the Parkes and Green Bank Telescopes by the Breakthrough Listen collaboration. Advances in detection technology have led to an exponential growth in data, necessitating innovative and efficient analysis methods. This problem is exacerbated by the large variety of possible forms an extraterrestrial signal might take, and the size of the multidimensional parameter space that must be searched. It is then made markedly worse by the fact that our best guess at the properties of such a signal is that it might resemble the signals emitted by human technology and communications, the main (yet diverse) contaminant in radio observations. We address this challenge by using a combination of simulations and machine learning methods for anomaly detection. We rank candidates by how unusual they are in frequency, and how persistent they are in time, by measuring the similarity between consecutive spectrograms of the same star. We validate that our filters significantly improve the quality of the candidates that are selected for human vetting when compared to a random selection. Of the $\sim 10^{11}$ spectrograms that we analyzed, we visually inspected thousands of the most promising spectrograms, and thousands more for validation, about 20,000 in total, and report that no candidate survived basic scrutiny. 

\end{abstract}

\section{Introduction}\label{s:intro}

For over half a century the Search for signs of Extra Terrestrial Intelligence (SETI) has progressed in lock step  with our soaring technological capabilities. For example, the pioneering Project Ozma used the 85-foot Tatel Telescope at the Green Bank Observatory to observe two stars with a frequency range of hundreds of Hz \citep{Drake1961}, while the current Breakthrough Listen (BL) searches obtain  high spectral and temporal resolution time-series over tens of GHz for millions of stars \citep{MacMahon2018, 2021Daniel}. But with great data come great challenges, as it is becoming increasingly difficult to search these data effectively for potential candidates with traditional methods. 

While it is difficult to define what the signals we are after would look like, we can define what we would like to exclude from our search. At radio frequencies (RF), astrophysical phenomena produce broadband spectra, while human communication and technology dominate the narrow-band regime. Human signals are plentiful and very diverse in their properties, and are a major contaminant to most RF observations, where they are labeled as RFI -- Radio Frequency Interference. Therefore, a good candidate detection would be narrow-band yet excluded from being of terrestrial origin. 

A number of recent works addressed the identification of terrestrial or otherwise narrow-band signals \citep{Siemion2013, Pinchuk2019, Margot2020}. A major component in recent efforts is TurboSETI \citep{turboseti}, a code designed to search for narrow-band drifting signals in RF data \citep{Enriquez2017, Price2019, Gajjar2021, Franz2022}. While TurboSETI is capable and efficient, it is limited by its detection logic that is optimized for, and therefore biased towards,  streaks in time-frequency representations. It also suffers from some algorithmic imperfections --- in particular, inefficient detection of signals with high drift rates \citep{Choza2024,Margot2023}.

It is unsurprisingly difficult to design a search for a vaguely defined and likely very rare source in abundant and messy data. However, this is the exact purpose of Anomaly Detection algorithms \citep{Ruff_2021}. Broadly speaking, these are a class of machine learning or AI methods that learn from the data what constitutes a `normal' object based on their similarities or a notion of density in some abstract space, in order to identify instances that do not inhabit the same regions of parameter space. Anomaly detection has been applied extensively in essentially all fields, from finance \citep{Hilal2022}, through medical imaging \citep{Tschuchnig_2022}, to industrial applications \citep{electronics12183971}. In astrophysics specifically, it has found application in planet hunting \citep{Shallue2018}, spectroscopic studies of galaxies and stars \citep{Baron2017,reis18,reis21}, and transient searches \citep{Villar_2021}, to name just a few examples. Anomaly detection is typically an unsupervised learning approach, in that it does not rely on labeled samples on which to train, nor does it usually have a clear objective function to minimize. Instead one typically relies on domain knowledge and experimentation to find the most rewarding approach. 

A few searches for technosignatures used supervised machine learning approaches, where they simulate possible signals inserted into the real data and leverage the power of deep neural networks that are trained to recover them \citep{Cox2018,Harp2019,Brzycki2020,Pinchuk2022}. \citet{Zhang2019} used anomaly detection on a self-supervised generative model. \citet{Ma2023,Ma2024} used a semi supervised auto-encoder architecture. \citet{JacobsonBell2024} used clustering of signals detected by turboSETI to find morphological outliers. 

Machine learning seems to be a natural solution given the computational challenge. While these methods are not computationally cheap, they scale better with the sample size than most alternatives, especially for prediction. The amount of data we have requires very efficient methods. However, even given an efficient method, the conceptual challenge remains. There are many ways to be abnormal, many ways to define similarity, and no general objective function that one can minimize in the pursuit of anomalies. Anomaly detection is an ill-defined problem by definition, and SETI is the epitome of anomaly detection problems. 

Here we attempt a new approach that aims to minimize biases with signal properties. We develop and apply a series of filters that reduce by 7 orders of magnitude the number of candidates that need human vetting. The data we use are described in Section \ref{s:data}. Section \ref{s:methods} details the various filters we develop and apply. In Section \ref{s:results} we validate our methods and visually rule out about 20 thousand potential candidates.

\section{Data} \label{s:data}
We use radio observations of nearby stars, obtained by the Breakthrough Listen initiative between the years 2016 and 2018 with the Parkes and Green Bank Telescopes (GBT). The observations were carried out in three bands that together cover the frequency range 1.10--3.45\,GHz (see Table \ref{t:telescopes}). We processed the data as described by \citet{Lebofsky2019}, converting raw voltage at the detectors into spectrograms.

In order to help separate the main contaminants, RFI, from a potential detection, each target star (`A') was observed in an alternating cadence with other pointings at similar coordinates (`B',`C', or `D'), over 6 consecutive 5-minute observations, forming a cadence like `ABABAB' (at Parkes) or `ABACAD' (at GBT). For the `ABABAB' strategy, the `B' consists of an off-source pointing at a constant 0\textdegree.5 offset in declination from the primary source, which is about 3 Full Width Half Maximum (FWHM) beam-widths at the lowest frequency.  In the `ABACAD' cadence the off-source targets are stars drawn from the Hipparcos catalog, with a distance of 1.2\textdegree--3.6\textdegree\ from the target, which ensures a separation of more than 8 FWHM beam-widths. The temporal resolution is about $20\,s$, with 16 time elements over the 5\,min exposure. 

Our sample initially consists of all the stars in \citet{Price2019}, a total of 1305  main-sequence stars, all within 50\,pc. However, to simplify our analysis, we restrict ourselves to observations with the same observed central frequency of 1475, 2300 and 3093--3094\,MHz in the S, L and 10-cm bands respectively. This selection preserves about 800 stars in each band. Similarly to \citet{Enriquez2017}, we exclude from our analysis any data between 1.2 and $1.33\,$GHz, as well as the $2.3-2.36\,$GHz range, which are attenuated by notch filters on the instrument due to high levels of RFI.

\begin{table}
    \centering
    	\caption{Telescopes used.}
        \begin{tabular}{c c c c c}
        \multicolumn{5}{c}{Table 1}                                                                                                                     \\
        \multicolumn{5}{c}{Details of telescopes used in the search}                                                                                    \\ \hline
        Telescope  & Receiver & \begin{tabular}[c]{@{}c@{}}Frequency\\ (GHz)\end{tabular} & N$_{obs}$ & \begin{tabular}[c]{@{}c@{}}Central\\ frequency (GHz)\end{tabular} \\ \hline
        Green Bank & L band  & 1.10--1.90                                                 & 836     & 1.475                                                  \\
        Green Bank & S band  & 1.80--2.80                                                 & 819     & 2.300                                                  \\
        Parkes      & 10 cm   & 2.60--3.45                                                 & 796     & 3.094                                                  \\ \hline
        \end{tabular}
    \label{t:telescopes}
\end{table}

Following \citet{Siemion2011}, we normalize and clean the spectrograms. First, every flux value is divided by the mean at that time over the observing window. Next, we remove DC noise spikes following \citet{Lebofsky2019}, by replacing them with the average value at neighboring frequencies. Last, to account for sensitivity variations across the band, we divide each channel by a B-spline interpolation over the integrated band-pass. We use the SCIKIT-LEARN \citep{Boor2001} implementation of the B-spline interpolation.

Like most searches for technosignatures, we focus here on narrow-band candidates ($\sim$Hz), where we expect engineered signals to dominate greatly over natural phenomena (see for example  \citealt{Siemion2013}). This choice allows us to cut the spectrograms into more manageable pieces. Specifically, we cut down the spectrograms to images with a width of 80 pixels in the frequency direction, or about 220\,Hz, with an overlap of 5 pixels between adjacent frequency windows, in order to make sure that no candidate is missed due to being cut at the edge. As a result, every observed cadence is composed of 6 pointings per cadence (`A', `B', etc.), times 16 time elements, times 80 frequency elements, and we have about $10^7$ of these to search for potential signals, per star-band combination, a staggering $\sim 10^{11}$ spectrograms in total.

\section{Methods}\label{s:methods}
Our method\footnote{In the 1997 movie `Contact' Jodie Foster's character Ellie Arroway is shown wearing large  headphones, listening to signals as they are obtained by the radio telescope. As a tribute to this endearing but silly moment of movie science, we call our pipeline LE2.0.} is composed of a series of filters which we apply to reduce the number of potential candidates from $\sim 10^{11}$ all the way to a few thousand that undergo human inspection. Due to the size of the dataset, the first step is the most onerous, so we intentionally keep it computationally simple. Using cross-correlation and dimensionality reduction, calibrated by simulations, we reduce the number of candidates by five orders of magnitude to $\sim 10^{6}$ (Section \ref{s:xcorr}). We then compute two quality metrics to rank candidates: a frequency score (Section \ref{s:freq}), and a similarity score (Section \ref{s:sim}). Simply put, the frequency score ranks higher those candidates that appear in quieter regions of the data, where they are less likely to be RFI. The similarity score is designed to prefer candidates that show a more consistent signal morphology as time progresses. These are detailed below.

\subsection{Cross-correlation filter} \label{s:xcorr}

\begin{figure*}
	\includegraphics[width=\textwidth]{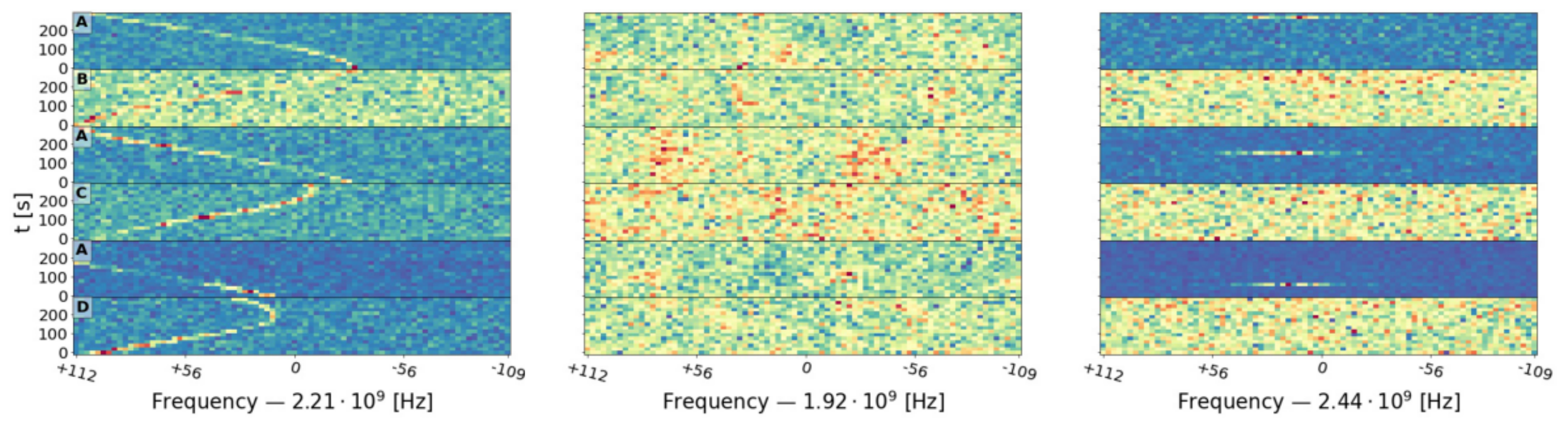}
    \caption{Three examples of observed cadences of the star HIP\,66193. The spectrograms on the left show loud RFI throughout the cadence, the middle example shows essentially background noise, and on the right we show what could be a viable candidate based on this view, having a detectable signal only on target.}
    \label{f:example_sequences}
\end{figure*}

We wish to perform a search that is minimally biased, as we have no real preconceived notion of the signal properties, other than its narrow frequency range, that we already assumed and justified in Section~\ref{s:intro}. In this parameter space our main (and effectively only) contaminants are RFIs. The only discernible feature that would allow us to separate a potential candidate from its earthly counterparts is its (dis)appearance in the alternating observing cadence. A good candidate appears only in the `A' on-target observations, and not in the `BCD' observations. A few representative examples are shown in Figure \ref{f:example_sequences}. The cadence in the leftmost panel is dominated by obvious RFI in all its panels, the central cadence is dominated by background noise, and the rightmost seems like a viable candidate, having detectable signal only on target. 

\begin{figure}
	\includegraphics[width=\columnwidth]{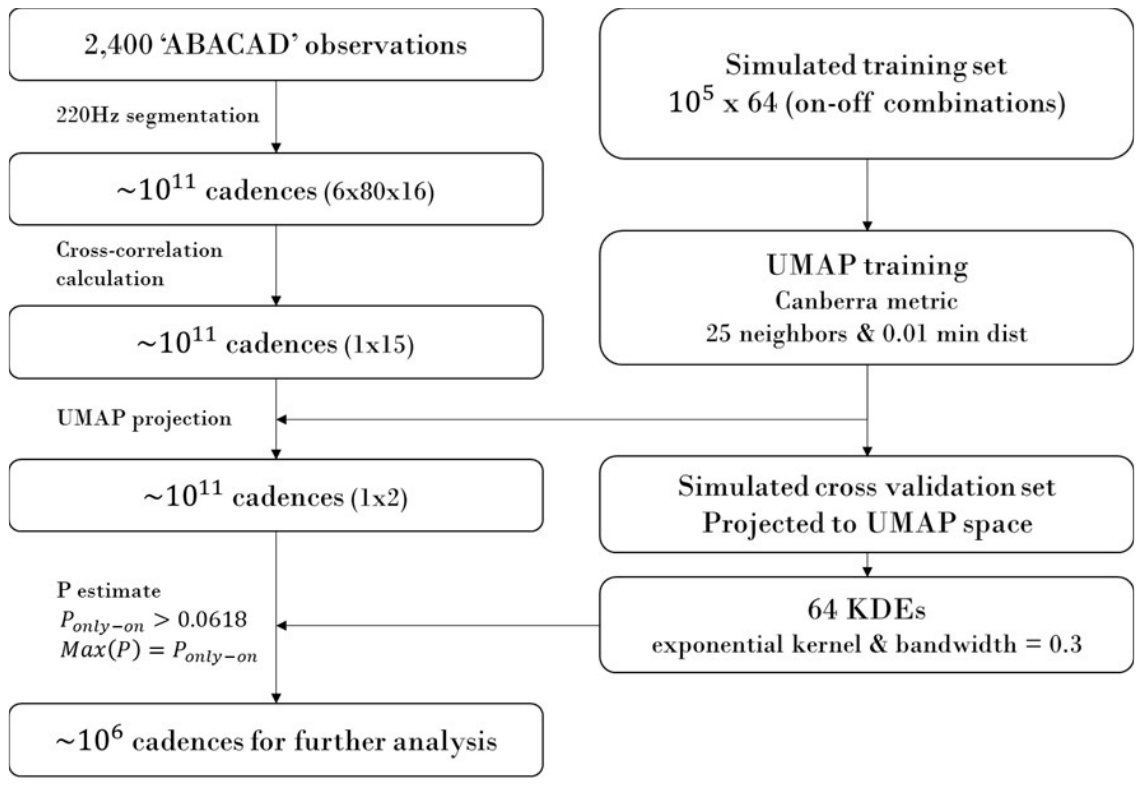}
    \caption{Schematic flow chart describing our first filter. We use dimensionality reduction and simulations to identify sources that have a detectable signal on target -- in the `A' parts -- but not in the alternates, which allows us to reduce the number of candidates by 5 orders of magnitude.}
    \label{f:schema}
\end{figure}

We have of order $10^{11}$ cadences of 6 spectrograms, 16 by 80 pixels each. Searching for patterns in that amount of data is a computational challenge with any algorithm. The steps we describe here, of cross-correlation and remapping to 2D, were the most computationally intensive stages of analysis. Overall, the analysis took hundreds of thousands of CPU hours, though this includes downloading and pre-processing the data. A schematic diagram detailing the steps we perform for the cross correlation filtering is shown in Figure \ref{f:schema}. 

In order to reduce the complexity of the calculation we rely on feature extraction to lower the dimensionality of the original data, and on simple operations that allow us to focus on the most promising candidates. For every observing cadence, we calculate the cross-correlation (CC) between each pair of spectrograms within the cadence (`A' with `B',`A' with `C', etc.), and tabulate the maximum, minimum, mean, median, and standard deviation values of the CC matrix for each pair. This results in 15 unique features (the upper non-diagonal elements of a 6 by 6 matrix) per cadence, reduced from the original 7680 pixels. These features should allow us to separate the cadences we are after from the rest.

We then map these 15 dimensions further down to two dimensions using UMAP \citep{McInnes2018}. UMAP is a dimensionality reduction algorithm, very similar to t-SNE \citep{Maaten&Hinton2008} that  maps the data to lower dimensions while conserving as much as possible the distances between neighbors. This comes at the inevitable price of distorting larger distances, but in doing so, the algorithm tries to preserve the underlying manifold (assumed or expected to be of low dimensionality) over which the data are distributed. Two key advantages of UMAP over similar methods are that it is relatively fast, and that it can be trained on one dataset, and produce a mapping from the original space to the UMAP plane that can be used to map other data that it has not seen before. At the end of this process every observing cadence of a star in a given band has been reduced to just two numbers -- XY coordinates in the abstract plane created by UMAP. 

In order to identify only those cadences that have a detectable signal on target -- in the `A' parts -- but not in the alternates, we use simulations. We simulate various signals, in various parts of the cadence, to find where these map to in the UMAP plane. We use the Setigen package \citep{Brzycki2022} to synthesize the signals, and the S-band cadence of HIP\,17147 for the  background noise, though our results are not sensitive to this choice. We simulate signals with all the time and frequency profiles that are available in the package, and with an amplitude that is drawn uniformly between zero and 4 times the maximum of the noise in the frequency window. Each of the 6 panels in a cadence can either have a signal or not, resulting in 64 options. For each of these options we generate 20,000 signals, for a total of 1,280,000 signals, half of which are used for training, and half for validation. 

UMAP has a few hyper-parameters that need to be set. Two of the most significant ones are \textit{n neighbors} (number of nearest neighbors; used to balance local vs global structure), and \textit{min dist} (minimum distance; sets how tightly packed are the points in the reduced space), for which we chose the values 25 and 0.01, respectively. These resulted in qualitatively good looking clusters (i.e., relatively well-defined and separated clusters), although our experimentation indicates that our results are not sensitive to these choices. A more critical choice is the distance metric used between the points, which does not have to be Euclidean. Here we found better results when using the Canberra metric which is a weighted version of the Manhattan distance measure.

\begin{figure*}
	\includegraphics[width=\textwidth]{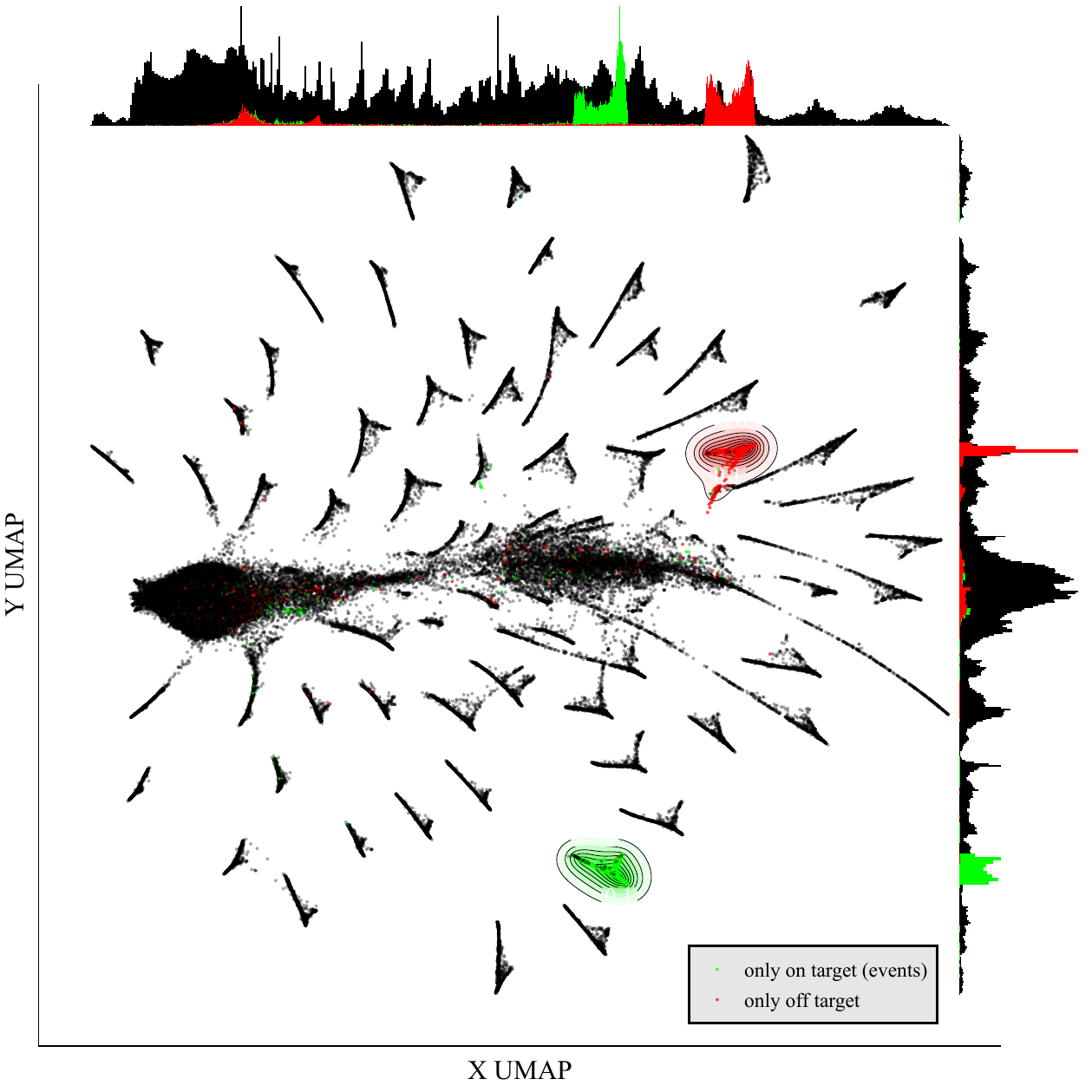}
    \caption{Mapping of simulated signals onto an abstract UMAP space. Every point is a simulated signal  from our validation dataset in any or all of the 6 windows of a cadence. In green we mark the signals that are only on-target, while in red we show those that exclusively off target for comparison. All the other clusters in black are various on-off combinations. In light green (red) we delineate the probability derived from the exclusively on(off)-target categories of the training set. In the margins we show the marginal distribution of the validation set. In addition to the well defined clusters, one can see a large group near the left of the plane. Visual inspection shows that this is where undetected weak signals, i.e., background noise, cluster.}
    \label{f:umap64}
\end{figure*}

In Figure \ref{f:umap64} we plot the UMAP mapping of the simulated signals from the validation set in black, with the marginal distributions to help visualize the relative densities of points. As can be seen, the simulated signals cluster clearly into groups, and more importantly, the simulated signals that are exclusively on-target, which we plot in green, define a well-separated region in the UMAP space. We also show for example that the only-off group (where the signal is in all panels \textit{except} the target) behaves in a similar fashion. 
To define and measure the `only on-target' region, we use the distribution of the simulated samples as measured by kernel density estimation (KDE) for each of the 64 categories in the training set. We use the \textit{SCIKIT-LEARN} \citep{Pedregosa2011} implementation of KDE, with an exponential kernel and a value of 0.3 for the bandwidth parameter, which seemed to recover well the distribution of the good (only on-target) candidates in the validation set as well. This gives us a probability that a cadence belongs to each of the 64 categories, based solely on its UMAP coordinates. While the probabilities do not add up to unity, the clusters are well separated, and as we show below, we can use these numbers to get a useful separation. The validation set maps consistently to the regions as defined with KDE, based on the training set, with the on-target region contaminated by a negligible fraction of $\sim 10^{-5}$ interlopers.

We examine by eye cadences that ended up in the large cluster near the center-left of the UMAP space, and we find that it is populated by low signal to noise ratio (SNR) cadences, or simply background noise. This explains why this cluster aggregates cadences from all the categories, whenever the simulated signal is too weak to be detected. The second grouping in the center-right is populated by always-on simulations.

\begin{figure}
	\includegraphics[width=\columnwidth]{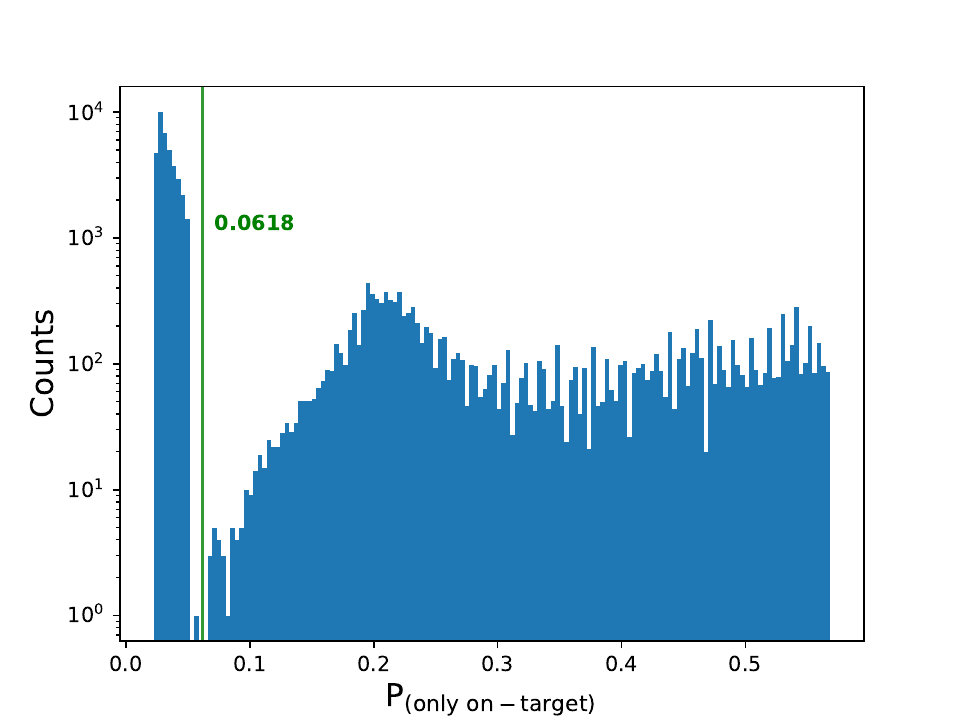}
    \caption{Histogram of the probability that a candidate is in the `only on-target' category, using 20 stars. There is an obvious cut-off near 0.0618 which we use as a lower bound to select candidates for further analysis.}
    \label{f:kde_hist} 
\end{figure}

We choose for further analysis only candidates for which the highest of the 64 probabilities is in the `only on-target'
category, and only if this probability is higher than 0.0618. This somewhat arbitrary threshold (which is highly dependent on our choice for the bandwidth), was determined by looking at the distribution of probabilities for a representative sample of candidates from 20 real observed targets. As can be seen in Figure \ref{f:kde_hist}, there is a natural cutoff near that value for our choices of hyper-parameters. This step reduces the number of candidates by a factor of $\sim 10^{5}$, leaving $\sim 10^{6}$ for further analysis, from the original $\sim 10^{11}$ cadences. This is still orders of magnitudes too many for human vetting, so we proceed to develop and apply finer filters.

\subsection{Frequency score}\label{s:freq}

\begin{figure*}
	\includegraphics[width=\textwidth]{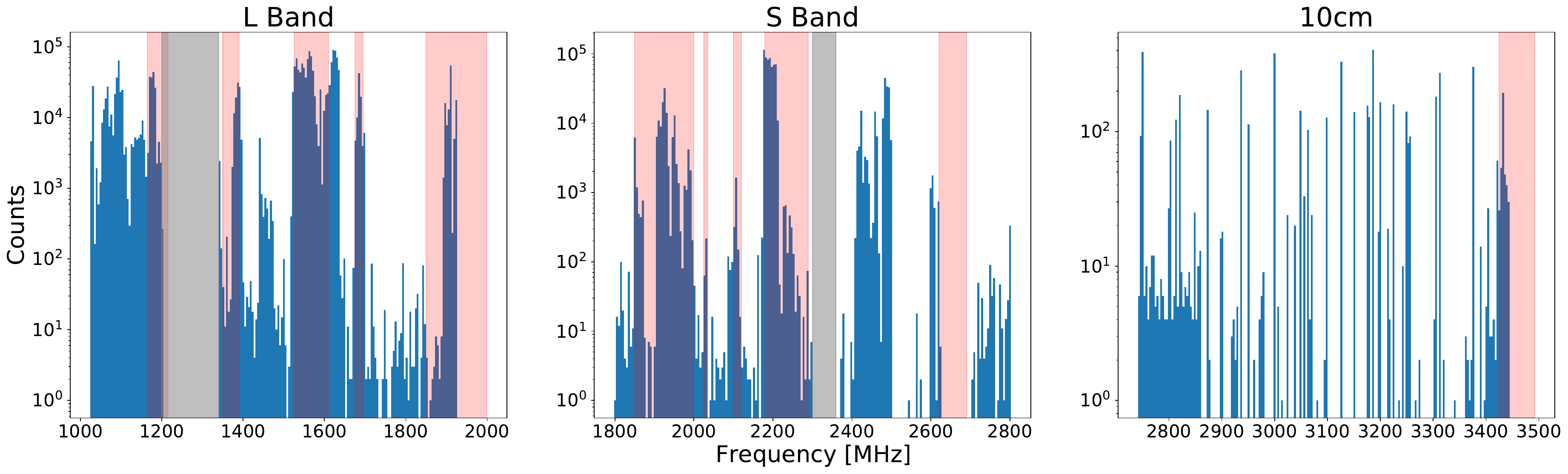}
    \caption{Histogram of all the candidates in each band, after the cross-correlation filter. They seem to prefer some frequencies to others, in a manner that is somewhat correlated with pink bands that are known to be heavily contaminated by RFI,  from \citet{Price2019}. The grey bands were completely excluded from our analysis.}
    \label{f:freq_hist}
\end{figure*}

In Figure \ref{f:freq_hist} we show the distribution of the number of candidates per star-band combination. Unsurprisingly, RFI occurs preferentially at some frequencies, though our busiest frequencies only partially overlap with the bands with intense human activity, as collated by \citet{Price2019}, which we mark in pink. Our search is limited by the number of false positives we are capable of examining manually\footnote{This would not be true for an airport suitcase scanner, for example, that is limited by the number of false negatives it should let through -- none -- at the inevitable cost of as many false positives as necessary.}. It therefore makes more sense to concentrate the effort, and search where the contamination from RFI is minimal. We do so by rejecting specific RFI-heavy observations, as well as similarly noisy frequency ranges. 

Since the matching between our candidate list and the known frequency windows is only indicative, we do not use them directly. Instead we rank a candidate's potential as being real, or effectively, its relative priority for human vetting, as inversely proportional to the signal density in our data in its frequency window. Since 65\% of our candidates come from just 5\% of the star-band combinations, we  discard these candidates. While there is some chance we would reject a loud but real source, we consider it unlikely, and a risk worth taking given our constraints.

We want to find outliers to the distribution of RFI, so we would like to know the probability density as a function of frequency at the candidate's position, without using said candidate to derive that probability. This means we would need to fit as many models as candidates,  every time fitting all the candidates except the one in question. This would be computationally challenging if not impossible with any model. Instead we make use of random sampling and bagging to obtain models that are less sensitive to outliers. We determine the density using a Gaussian Mixture Model, where the number of Gaussians we fit is a hyper-parameter we discuss below. In every band we sample X candidates Y times and derive Y probability density functions which we then average. 
The candidates are then assigned a score which is the inverse of the mean probability density at that frequency, re-scaled to be between 0 and 1, such that better candidates have higher scores. This allows us to easily prioritize candidates in quiet areas of the spectrum. 

\begin{figure*}
	\includegraphics[width=\textwidth]{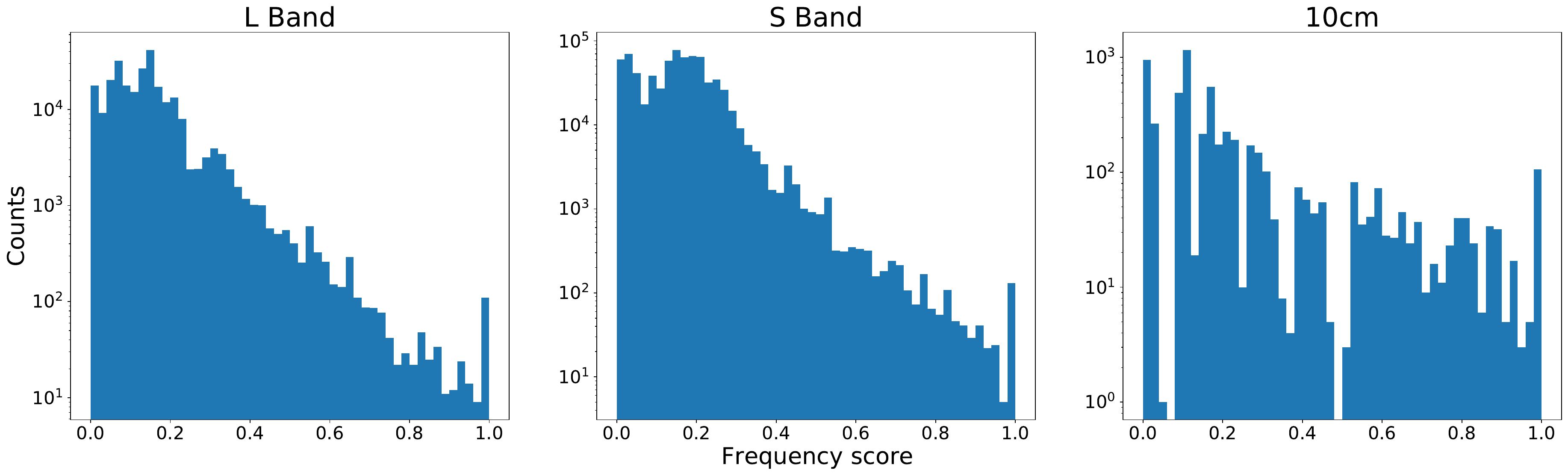}
    \caption{Histogram of the frequency score of all candidates in each band, after the cross-correlation filter. We use these scores to rank candidates for human vetting. We examined and ruled out all the candidates in the top one percent of each band.}
    \label{f:freq_scores}
\end{figure*}

We determine the number of GMM components in each band by gradually increasing it, until the scores for outliers stop depending on the number of Gaussians. If that number is too small the existing Gaussians are `inflated' by the fit in order to better represent more points. This leads to a multi-modal distribution of frequency scores. When we use 200, 200 and 50 components for the S, L and C band, respectively, we obtain the distributions shown in Figure \ref{f:freq_scores}.

\subsection{Similarity score}\label{s:sim}

For our next and last filter we return to the actual spectrograms which we now want to prioritize by similarity. We expect a real ET signal to be similar to itself in all the `A' observations, and be dissimilar from the off-target observations. For computational reasons, we first linearly reduce the resolution of the spectrograms, by a factor of 5 in frequency, and 2 in time, bringing them to a more manageable size of $16\times8\,$pixels each, from the $80\times16\,$pixels before. 

We then train UMAP, using 20\% of the candidates chosen from random cadences, a Euclidean metric, \textit{n\_neighbors}=25, and \textit{min\_dist}=0.01, which gives qualitatively satisfactory clustering. We compute the similarity score of a candidate by looking at the distances in UMAP space between the on- and off-target observations of the candidate cadence. A good candidate will have all the `on-target' observations similar to each other, and therefore they will be close to each other in the UMAP space, relative to the distances to the `off-target'. This ratio is the similarity score for a candidate, and can be written as:
\begin{equation} 
   \textrm{Similarity Score}=1-\sum_{i \in \textrm{on}} \frac{ \sum_{j>i,\, j \in \textrm{on}} \sqrt{(x_i - x_j)^2 + (y_i - y_j)^2 }}{\sum_{k \in \textrm{off}} \sqrt{(x_i - x_k)^2 + (y_i - y_k)^2}}, 
\end{equation} 
 where the coordinates are in the UMAP plane, and the indices refer to the 6 spectrograms in a cadence that are either on- or off-target. 

\begin{figure*}
	\includegraphics[width=\textwidth]{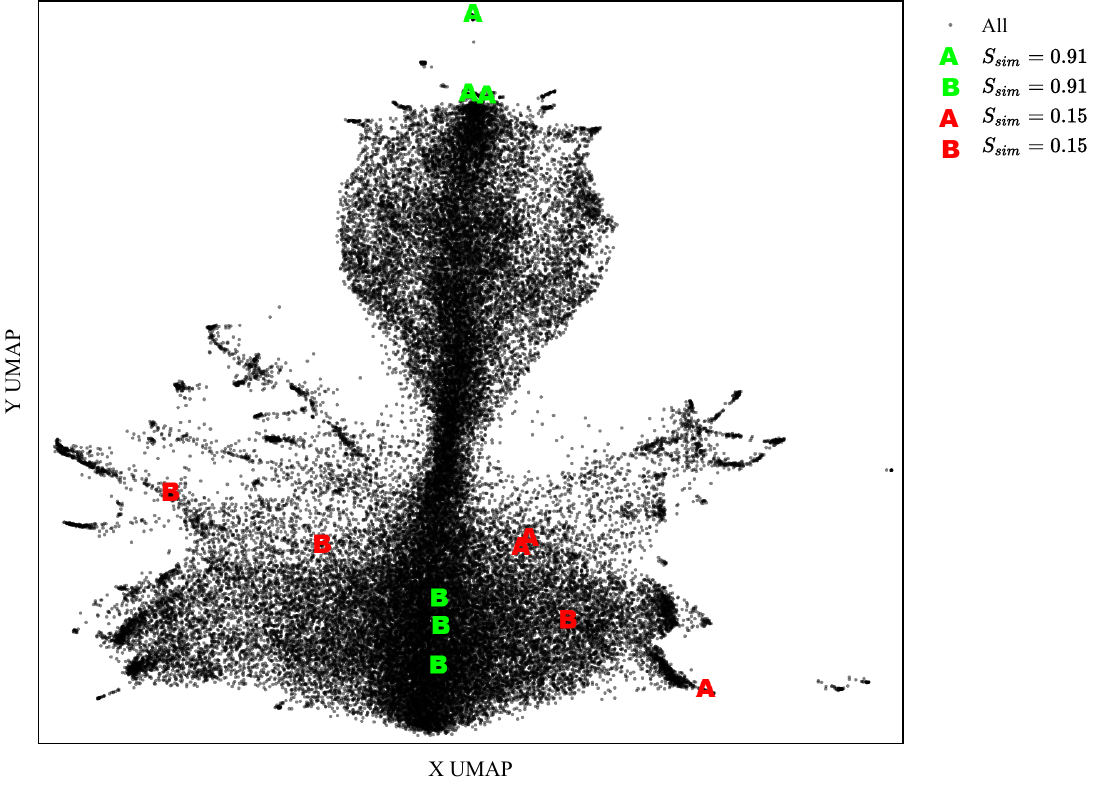}
    \caption{Mapping of individual observations of the entire candidates list onto an abstract UMAP space. In green we show a promising candidate, where all the `A' observations seem similar to each other, and different from the off-target `B' observations, which is reflected in a high similarity score, while the candidate in red shows no such clustering.}
    \label{f:sim_umap}
\end{figure*}

We scaled the similarity scores to range from 0 to 1, with higher scores assigned to better candidates. Figure \ref{f:sim_umap} shows the distribution of the spectrograms in UMAP. We mark two representative examples. The example in green, has its `A' and `B' observations grouped separately, giving it a high similarity score, near 1,  while the red candidate is dispersed and gets a low score.

\section{Human vetting and Results} \label{s:results}

Vetting was primarily done by the lead author, and is composed of two stages. In a first rapid stage, the candidate 6-spectrogram cadence is examined by eye, exactly as it is fed to the machine. About $90$\% are typically discarded at this stage, usually as RFI, showing signals in more than just the `on-target' panels. We call the candidates that clear this first stage `top-candidates', and they require a slightly more thorough examination. A good example of a top candidate is shown in the right panel of Figure  \ref{f:example_sequences}. In the second stage, top-candidates are inspected more carefully, by `zooming out' in time and in frequency, and by examining data surrounding the candidate, both before and after normalization. No candidate required further scrutiny to be ruled out, they were all easy to unambiguously reject. 

We can use the relative number of top-candidates in a sub-sample as an indication for the sample's purity or quality, a measure of the usefulness of our scores. A perfect system would supply us with top-candidates exclusively, and that ratio would be 1. We construct 4 sub-samples. We randomly select over 300 stars/band combinations that have at least  1000 candidates each (some have none or few). The first sample is composed of the top one percentile in frequency score for each star-band selected. The second sample is composed of the top one percentile in similarity score for each star-band selected. The third, which should be the most promising for the actual search, is composed of the top 1000 candidates when considering both scores equally. The 4th sample of 3000 candidates, is a control and is completely random and independent of either score. These number were selected to produce a few thousand candidates per sample. 

The frequency-score-based sample had about 7 percent of top candidates, the similarity-only sample 14 percent, and when combining both we reach about 22 percent. All of these are significantly higher than the randomly-selected control sample, which had a success rate of less than 3 percent. This shows that our scores both contribute to selecting better candidates, by a significant margin. Of the nearly 20k candidates that we scanned, a total of about 2000 required a second stage. No candidate survived this additional scrutiny, and all were ruled out as RFI or other instrumental noise. We did not find any viable ET candidate. 

Our overall approach required quite a few parameters to tune and many decisions to make, despite our best efforts to be minimally biased. This is an unavoidable and tricky aspect of discovery with anomaly detection methods. This is much less problematic when one discovers something, but with non-detection it is challenging to make a quantitative statement. Any simulation will be obviously biased by the populations we choose to simulate, which is exactly what we were trying to avoid as much as possible. When one finds something, the proof is proverbially in the pudding. 

We can however estimate the SNR per pixel of a candidate by comparing the pixel value to the median absolute deviation in a given spectrogram. But to know the SNR of the candidate we need to identify which pixels contribute to the signal, and which are mostly background, which cannot be done simply or automatically in an unbiased way. Instead, we define $X10$, the number of pixels in a spectrogram with SNR greater than 10. From simulation, where we know the true SNR, we find that this value is a reasonable tracer of the total SNR of a candidate. The candidates that were chosen for human vetting had an $X10$ in the range of 1--300, where about 10 percent are above 100. Similarly to \citep{Price2019} our S-band candidates had the highest relative SNR, and 10 cm had the lowest. 

Another telling comparison is to the findings of \citet{Ma2023} who used a deep learning approach on a similar sample to ours. \citet{Ma2023} focused on the L-band and used much larger frequency windows of $11\,$kHz (compared to ours of $220\,$Hz). As a result, they were sensitive to higher drift rates. Nevertheless, we find that our candidates and theirs have very similar frequency distributions in that band. 

In table \ref{t:ma_events} we show the 8 most promising candidates that they find. Our cross correlation filter did not pass only 2 of their candidates, $MLc1$ and $MLc2$, who also have the largest drift rates and lowest SNR of their sample. 3 more candidates ($MLc4$ to $MLc6$) were filtered out because they were in the 5\% of the star-band combinations with the highest background. The last 3, $MLc3$, $MLc7$, and $MLc8$ were not selected for human vetting either, though $Mlc8$ came very close. Looking at the frequency scores for these last 3, as well as the similarity score for the 6 candidates that passed the first filter, it is reassuring to see that all of them are ranked very highly, most of them in the top few percent.

\begin{table}[ht]
    \centering
    \caption{Promising candidates from \citet{Ma2023}}
    \begin{tabular}{l l c c c c}
        \hline
        ID   & Target       & Drift rate [Hz/s] & SNR & Frequency rank                 & Similarity rank       \\ \hline
        MLc1 & HIP 13402    & +1.11(25) & 7   & -                              & -              \\
        MLc2 & HIP 118212   & -0.44(7)  & 16  & -                              & -              \\
        MLc3 & HIP 62207    & -0.05(10) & 58  & 0.980                      & 0.967      \\
        MLc4 & HIP 54677    & -0.11(3)  & 30  & noisy star-band                & 0.934       \\
        MLc5 & HIP 54677    & -0.11(2)  & 45  & noisy star-band                & 0.989      \\
        MLc6 & HIP 56802    & -0.18(4)  & 40  & noisy star-band                & 0.890      \\
        MLc7 & HIP 13402    & +0.10(2)  & 129 & 0.989                    & 0.983      \\
        MLc8 & HIP 62207    & -0.126(10)& 34  & 0.998                      & 0.978      \\
        \hline
    \end{tabular}
    \label{t:ma_events}
\end{table}

The two most onerous parts of our pipeline are the first and the last steps. The first, the cross correlation filter, which we ran on $\sim10^{11}$ spectrogram, had to be as simple as possible, and easy to parallelize, in order to be executed within a reasonable wall-time. It still took many thousands of CPU hours, dwarfing all other steps combined. The last step, the manual vetting of candidates, is taxing for the human in the loop. In future works, one approach we would like to experiment with, is a multi-scale system, that does not tie us to a specific size in time-frequency space. As discussed, `zooming-out' was the most effective way to rule out candidates. If these were automatically rejected we would be able to scan many more potential candidates.

\section{Conclusions}
SETI exemplifies the challenge of identifying new phenomena in vast, noisy datasets. We leveraged a minimal set of assumptions and simple operations, in order to search efficiently for technosignatures in radio observations of nearby stars. Instead of relying on signal identification, that requires one to define the signal that is sought (e.g., streaks in time-frequency, as done with TurboSETI), we used cross-correlation, dimensionality reduction, and statistics on their occurrence as a function of frequency, to separate RFI from potential candidates. Of the $10^{11}$ spectrograms of more than 300 nearby stars that we analyzed, we visually inspected about 20,000 of the most promising candidates and ruled them all out. Future work should explore incorporating adaptive multi-scale techniques to further automate the vetting process and expand search efficiency.

\section*{Acknowledgments}

This research was funded in part by the Koret Foundation, the Kavli Institute for Particle Astrophysics and Cosmology at Stanford University, and by grant NSF PHY-2309135 to the Kavli Institute for Theoretical Physics (KITP). S.P. and D.P. acknowledge support from Israel Science Foundation (ISF) grant 541/17, and by grant 2018017 from the United States-Israel Binational Science Foundation (BSF). S.C., A.P.V.S., and M.L. acknowledge support from the Breakthrough Initiatives. The Breakthrough Prize Foundation funds the Breakthrough Initiatives, which manages Breakthrough Listen. The Green Bank Observatory facility is supported by the National Science Foundation, and is operated by Associated Universities, Inc., under a cooperative agreement.

\bibliography{library}{}
\bibliographystyle{aasjournal}

\end{document}